\documentclass[twocolumn,prl,showpacs]{revtex4}%
\usepackage{graphicx}%
\usepackage{amsmath}%
\usepackage{amsfonts}%
\usepackage{amssymb}
\usepackage{bm}

\def\s{{\sigma}}
\def\e{{\epsilon}}
\def\k{{ {\bm k} }}

\def\q{{ {\bm q} }}
\def\Q{{ {\bm Q} }}

\def\w{{\omega}}
\def\a{{\alpha}}

\def\g{{\gamma}}

\begin{document}
\title{Structure of Neutron-Scattering Peak in both $s_{++}$-wave and
$s_\pm$-wave states \\ of an Iron pnictide Superconductor
}
\author{Seiichiro \textsc{Onari}$^{1}$,
 Hiroshi \textsc{Kontani}$^{2}$, and Masatoshi \textsc{Sato}$^2$}
\date{\today }

\begin{abstract}
We study the neutron scattering spectrum in iron pnictides 
based on the random-phase approximation in the five-orbital model,
for fully-gapped $s$-wave states
with sign reversal ($s_\pm$) and without sign reversal ($s_{++}$).
In the $s_{++}$-wave state, we find that a prominent hump structure 
appears just above the spectral gap,
by taking account of the quasiparticle damping $\gamma$
due to strong electron-electron correlation:
As the superconductivity develops, the reduction in $\gamma$
gives rise to the large overshoot in the spectrum above the gap.
The obtained hump structure looks similar to the 
resonance peak in the $s_\pm$-wave state, 
although the height and weight of the peak 
in the latter state is much larger.
In the present study, experimentally observed broad spectral peak 
in iron pnictides is naturally reproduced by assuming the $s_{++}$-wave state.
\end{abstract}

\address{
$^1$ Department of Applied Physics, Nagoya University and JST, TRIP, 
Furo-cho, Nagoya 464-8602, Japan. 
\\
$^2$ Department of Physics, Nagoya University and JST, TRIP, 
Furo-cho, Nagoya 464-8602, Japan. 
}
 

\sloppy

\maketitle


Since the discovery of superconductivity in iron pnictides with high
transition temperature 
$(T_c)$ next to high-$T_c$ cuprates\cite{Hosono},
the structure of the superconducting (SC) gap 
has been studied very intensively.
The SC gap in many iron pnictides is fully-gapped and 
band-dependent, as shown by the penetration depth measurement \cite{Matsuda}
and the angle-resolved photoemission spectroscopy (ARPES)
 \cite{ARPES1,ARPES4},
except for P-doped Ba122 \cite{AsP}.
The fully-gapped state is also supported by the rapid suppression 
in $1/T_1$ ($\propto T^{n}$; $n\sim4-6$) below $T_{\rm c}$ 
\cite{Sato-T1,Mukuda,Grafe}.

In iron pnictides, the nesting of the Fermi surface (FS)
between hole- and electron-pockets is expected to induce
the antiferromagnetic (AF) fluctuations in doped metal compounds.
Since fully-gapped sign-reversing $s$-wave state ($s_\pm$-wave state)
is a natural candidate \cite{Kuroki,Mazin},
it is urgent to clarify the sign reversal in the SC gap
via phase-sensitive experiments.
One of the promising methods is the neutron scattering measurement:
Existence of the resonance peak at a nesting wavevector ${\bf Q}$
is a strong evidence for AF fluctuation mediated 
superconductors with sign reversal
\cite{pines,chubukov-resonance,takimoto-moriya}.
The resonance condition is $\w_{\rm res}<2\Delta$,
where $\w_{\rm res}$ is the resonance energy and 
$\Delta$ is magnitude of the SC gap at $T=0$.
The resonance peak has been observed in many
AF fluctuation mediated unconventional superconductors, like
high-$T_c$ cuprates \cite{iikubo-sato,ito-sato,keimer-highTc},
CeCoIn$_5$ \cite{stock-CeCoIn5}, and UPd$_2$Al$_3$ \cite{sato-UPd2Al3}.

Neutron scattering measurements for iron pnictides have been performed 
\cite{christianson,keimer,zhao,qiu}
after the theoretical predictions \cite{maier-scalapino,eremin}.
Although clear peak structure was observed in
FeSe$_{0.4}$Te$_{0.6}$ \cite{qiu} and 
BaFe$_{1.85}$Co$_{0.15}$As$_2$ \cite{keimer},
its weight is much smaller than that in high-$T_c$ cuprates and CeCoIn$_5$, 
and the resonance condition $\w_{\rm res}<2\Delta$
is not surely confirmed, as we will discuss later.

Nonmagnetic impurity effect 
also offers us useful phase-sensitive information.
Theoretically, $s_{\pm}$-wave state should be very fragile
against impurities due to the interband scattering \cite{onari-impurity};
the predicted critical residual resistivity $\rho_{\rm imp}^{\rm cr}$
for vanishing $T_{\rm c}$ is about $20\ \mu\Omega$cm.
However, experimental $\rho_{\rm imp}^{\rm cr}$ reaches 
$\sim750\ \mu\Omega$cm, which corresponds to the minimum metallic 
conductivity $4e^2/h$ per layer \cite{Sato-imp}.
Since this result supports a conventional $s$-wave state without sign 
reversal ($s_{++}$-wave state), we have to resolve the 
discrepancy between neutron scattering measurements
and the impurity effects.


In this letter, we study the dynamical spin susceptibility 
$\chi^{\rm s}(\w,\Q)$ based on the five-orbital model \cite{Kuroki}
for both $s_{++}$ and $s_\pm$ wave states,
and discuss by which pairing state the experimental results are reproducible.
In the normal state, $\chi^{\rm s}(\w,\Q)$ is strongly suppressed 
by the quasiparticle damping $\gamma$
due to strong correlation.
However, this suppression diminishes in the SC state since 
$\gamma$ is reduced as the SC gap opens.
For this reason, a prominent hump structure 
{\it unrelated to the resonance mechanism}
appears in $\chi^{\rm s}(\w,\Q)$ just above $2\Delta$
in the $s_{++}$-wave state.
In the $s_\pm$-wave state, very high and sharp resonance 
peak appears at $\w_{\rm res}<2\Delta$.
We demonstrate that the broad spectral peak observed in iron pnictides 
is naturally reproduced based on the $s_{++}$-wave state,
rather than the $s_\pm$-wave state.

\begin{figure}[!htb]
\includegraphics[width=\linewidth]{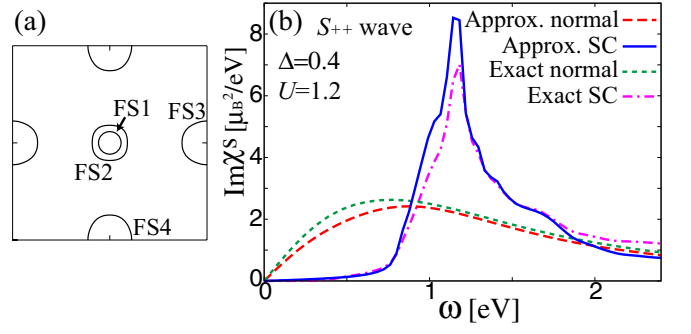}
\caption{
(Color online) 
(a) FSs in iron pnictides.
(b) $\w$-dependence of Im$\chi^s(\w,{\bm Q})$
for the $s_{++}$-wave state ($\Delta=0.4$) and the normal state, where the unit of energy is eV.
The ``exact result'' is obtained by eq. (\ref{chi-exact}),
and the ``approximate result'' is obtained by eq. (\ref{chi-approx}).
}
\label{Fig1}
\end{figure}

Now, we study the $10\times10$ Nambu BCS Hamiltonian ${\hat {\cal H}}_\k$
composed of the five-orbital tight-binding model and 
the band-diagonal SC gap introduced in ref. \cite{onari-impurity}.
The FSs are shown in Fig. \ref{Fig1} (a).
Then, the $10\times10$ Green function is given by
\begin{eqnarray}
{\hat {\cal G}}(i\w_n,\k) &\equiv&
\left(
\begin{array}{cc}
{\hat G}(i\w_n,\k) & {\hat F}(i\w_n,\k) \\
{\hat F}^\dagger(i\w_n,\k) & -{\hat G}(-i\w_n,\k) \\
\end{array}
\right)^{-1} 
 \nonumber \\
&=& (i\w_n{\hat 1}-{\hat \Sigma}_\k(i\w_n)-{\hat {\cal H}}_\k)^{-1} ,
 \label{eqn:G}
\end{eqnarray}
where $\w_n=\pi T(2n+1)$ is the fermion Matsubara frequency,
${\hat G}$ (${\hat F}$) is the $5\times5$ normal (anomalous)
Green function, and ${\hat \Sigma}_\k$ is the self-energy in the
$d$-orbital basis. 
For a while, we assume that the SC gap for the $\a$-th FS
is band-independent; $|\Delta_\a|=\Delta$.
Hereafter, the unit of energy is eV, unless otherwise noted.

Here, we have to obtain the spin susceptibility 
as function of real frequency.
For this purpose, it is rather easy to use the Matsubara frequency method
and the numerical analytic continuation (pade approximation).
In the present study, however,
we perform the analytical continuation 
before numerical calculation in order to obtain more reliable results.
The irreducible spin susceptibility in the singlet SC state is given by
\cite{takimoto-moriya}
\begin{eqnarray}
\hat{\chi}_{l_1l_2,l_3l_4}^{0{\rm R}}(\omega,\q)
&=&\frac{1}{N}\sum_\k\int\frac{dx}{2}
 \label{chi-exact}\\
& &\left[\tanh\frac{x}{2T}G^{\rm R}_{l_1l_3}(x_+,\k_+)
\rho_{l_4l_2}^{\rm G}(x,\k)\right.\nonumber\\
&+&\tanh\frac{x_+}{2T}\rho^{\rm G}_{l_1l_3}(x_+,\k_+)G^{\rm A}_{l_4l_2}(x,\k)\nonumber\\
&+&\tanh\frac{x}{2T}F^{\rm R}_{l_1l_4}(x_+,\k_+)
\rho^{{\rm F}\dagger}_{l_3l_2}(x,\k)\nonumber\\
&+&\left.\tanh\frac{x_+}{2T} \rho_{l_1l_4}^{\rm F}(x_+,\k_+)
F^{\dagger {\rm A}}_{l_3l_2}(x,\k)\right], 
\nonumber
\end{eqnarray}
where $x_+=x+\w$, $\k_+=\k+\q$, 
$l_i=1\sim5$ represents the $d$-orbital,
and A (R) represents the advanced (retarded) Green function.
$\rho_{ll'}^{\rm G}(x,\k)\equiv 
(G_{ll'}^{\rm A}(x,\k)-G_{ll'}^{\rm R}(x,\k))/(2\pi i)$
and 
$\rho_{ll'}^{\rm F(\dagger)}(x,\k)\equiv 
(F_{ll'}^{(\dagger)\rm A}(x,\k)-F_{ll'}^{(\dagger)\rm R}(x,\k))/(2\pi i)$
are one particle spectral functions.
Since $\rho_{ll'}^{\rm G,F}(x,\k)=0$ for $|x|<\Delta$,
Im$\hat{\chi}_{ll,l'l'}^{0{\rm R}}(\omega,\q)=0$ for $|\w|<2\Delta$.
That is, the particle-hole excitation gap is $2\Delta$.

Then, the spin susceptibility $\chi^s(\w,\q)$
is given by the multiorbital random-phase-approximation (RPA)
with the intraorbital Coulomb $U$, the interorbital Coulomb $U'$, 
the Hund coupling $J$, and the pair-hopping $J'$ \cite{Kuroki}:
\begin{equation}
\chi^s(\w,\q)=\sum_{i,j}\left[\frac{\hat{\chi}^{0{\rm R}}(\omega,\q)}
{1-{\hat S}^0\hat{\chi}^{0{\rm R}}(\omega,\q)}\right]_{ii,jj},
\label{RPA}
\end{equation}
where vertex of spin channel ${\hat S}^0_{l_1l_2,l_3l_4}=U$, $U'$, $J$ and $J'$ for
$l_1=l_2=l_3=l_4$, $l_1=l_3\ne l_2=l_4$ , $l_1=l_2\ne l_3=l_4$ and
$l_1=l_4\ne l_2=l_3$, respectively.
Hereafter, we put $J=J'=0.15$, $U'=U-2J$, $U=1\sim1.3$,
and fix the electron number as 6.1 (10\% electron-doped case).
In the present model, $\chi^s(0,\q)$ takes the maximum value when 
$\q$ is the nesting vector $\Q=(\pi,\pi/16)$. Due to the nesting,
$\chi^s(0,\Q)/\chi^0(0,\Q) \approx 1/(1-\a_{\rm St})$ is enhanced;
$\a_{\rm St}\ (\lesssim1)$ is 
the maximum eigenvalue of ${\hat S}^0\hat{\chi}^{0{\rm R}}(0,\Q)$
that is called the Stoner factor.

In strongly correlated systems, 
$\chi^s(\w,\q)$ is renormalized by the self-energy correction.
In nearly AF metals, for example, the temperature dependence of the self-energy
induces the Curie-Weiss behavior of $\chi^s(0,\Q)$.
At the moment, there is no experimental information on the $\k$-, $\e$-, 
and band-dependences of the self-energy.
Therefore, we phenomenologically introduce a band-diagonal self-energy as
${\hat \Sigma}_\k^{\rm R}(\e)= i\gamma(\e){\hat 1}$.
First, we estimate the value of $\gamma(\e)$ in the normal state.
Since the conductivity is given by $\s=e^2\sum_\a N_\a(0)v_{\a}^2/2\gamma(0)$,
where $N_\a(0)$ and $v_\a$ are the density of states (DOS) and
the Fermi velocity of the $\a$-th FS,
we obtain $\rho\approx(2\gamma {\rm [meV]})$ $\mu\Omega$cm
 \cite{onari-impurity}.
Since $\rho(T)-\rho(0)\sim (5 T[{\rm meV}])$ $\mu\Omega$cm 
in BaFe$_{1.84}$Co$_{0.16}$As$_2$ below 100 K \cite{rho-BaFeCoAs}, 
$\gamma(0)$ due to inelastic scattering is estimated as $2.5 T$
which is comparable to that in over-doped cuprates.
If we assume the relation $\gamma(\e)\propto (\pi T+\e)$ 
in nearly AF Fermi liquid \cite{Pines}, 
we obtain $\gamma(\e)\sim 2.5(T+\e/\pi)$.


Now, we calculate Im$\chi^s(\w,\Q)$ in both normal and 
$s_{++}$-wave SC states,
concentrating on the frequency $\w\sim2\Delta$.
To estimate the renormalization of Im$\chi^s(\w,\Q)$
due to the self-energy,
we have to know the value of $\gamma(\e)$ with $|\e|\sim\Delta$
in both normal and SC states.
Considering that $\gamma(\e)=2.5(T+\e/\pi)\sim 2\Delta$ 
at $T_{\rm c}=2.2$ meV and $\e=\Delta\sim5$ meV in BaFe$_{1.85}$Co$_{0.15}$As$_2$, 
in the  present study, we simply put $\gamma(\e)$ in the normal state 
at $T_{\rm c}$ as
\begin{eqnarray}
\gamma(\e)=\gamma_0
 \label{gamma0}
\end{eqnarray}
with $\gamma_0\gtrsim \Delta$.
In the present model, $\a_{\rm St}=0.84$ (0.79) for $U=1.3$ (1.2)
when $\gamma_0=0.1$ and $T=0.002$;
the $T$-dependence of $\a_{\rm St}$ is small when $\gamma_0$ is fixed.

In the SC state at $T\ll T_c$, $\gamma(\e)=0$ for $|\e|<3\Delta$ 
($=$ particle-hole excitation gap $2\Delta$
plus one-particle gap $\Delta$) \cite{chubukov-resonance},
and its functional form is approximately the same as that in the normal state
for $|\e|\gtrsim3\Delta$. Then, we put
\begin{eqnarray}
\gamma(\e)=a(\e)\gamma_s
 \label{a}
\end{eqnarray}
where (i) $a(\e)\ll1$ for $|\e|<3\Delta$,
(ii) $a(\e)=1$ for $|\e|>4\Delta$, and
(iii) linear extrapolation for $3\Delta<|\e|<4\Delta$.
We have confirmed that the obtained results are insensitive to the boundary of 
$|\e|$ ($4\Delta$ in the present case) between (ii) and (iii).
Although $\gamma_s$ at $T\ll T_{\rm c}$ should be smaller than 
$\gamma_0$ at $T=T_{\rm c}$, we simply put $\gamma_s=\gamma_0$
hereafter, which causes underestimation of the peak height of Im$\chi^s$.

Figure \ref{Fig1} shows Im$\chi^s(\w,{\bm Q})$
obtained by eqs. (\ref{chi-exact}) and (\ref{RPA})
for $U=1.2$, $\gamma_0=0.4$ and $T=0.01$.
In the $s_{++}$-wave SC state, we put $\Delta=\gamma_0$ and $a(3\Delta)=0.05$.
In calculating eq. (\ref{chi-exact}),
we use $256\times256$ $\k$-meshes, and 1000 $x$-meshes.
Although values of $\Delta$ and $\gamma$ in Fig. \ref{Fig1}
are very large to obtain enough numerical accuracy,
the ratio $\gamma_0/\Delta\sim1$ is consistent with experiments.
In the normal state, Im$\chi^s(\w,{\bm Q})$ is suppressed by large 
quasiparticle damping $\gamma_0\sim\Delta$.
In the SC state, the gap in Im$\chi^s(\w,{\bm Q})$ is $2\Delta$.
Since the particle or hole with energy $|\e|<3\Delta$
is free from inelastic scattering in the SC state, the lifetime of 
particle-hole excitation with energy $|\e|<4\Delta$ becomes long.
For this reason, Im$\chi^s(\w,\q)$ shows a large hump structure
for $2\Delta\lesssim\w\lesssim4\Delta$ below $T_{\rm c}$ in
$s_{++}$-wave state.


Unfortunately, we cannot put smaller $\Delta$ and $\gamma$  
in calculating eq. (\ref{chi-exact}) in the five-orbital model,
because of the computation time.
To solve this problem, we perform the $x$-integration in 
eq. (\ref{chi-exact}) approximately as follows:
When ${\hat \gamma}= \gamma{\hat 1}$,
the retarded (advanced) $10\times10$ Green function is expressed as
${\hat {\cal G}}_{m,m'}^{\rm R(A)}(x,\k)
= \sum_\a U_\k^{m,\a} (x+(-)i\gamma-E_\k^\a)^{-1} {U_\k^{m',\a}}^*$, 
where $E_\k^\a$ ($\a=1\sim10$) is the eigenvalue of ${\hat{\cal H}}_\k$
and ${\hat U}_\k$ is the corresponding $10\times10$ unitary matrix.
We promise that $E_\k^\a=-E_\k^{\a+5}$ for $1\le\a\le5$.
When $\gamma$ is sufficiently small, 
then $\rho_{ll'}^{\rm G(F)}(x,\k)\approx \sum_\a U_\k^{l,\a}
\delta(x-E_\k^\a){U_\k^{l'(+5),\a}}^*$, and thus eq. (\ref{chi-exact}) becomes 
\begin{eqnarray}
\hat{\chi}_{l_1l_2,l_3l_4}^{0{\rm R}}(\w,\q)\approx\frac{1}{N}
\sum_{k}\sum_{l,l'}\frac{f(E^l_\k)-f(E^{l'}_{\k+\q})}
{\w+E^{l}_\k-E^{l'}_{\k+\q}+i\Gamma_{ll',\k\q}}
\nonumber\\
\left[U^{l_1,l'}_{\k+\q}{U^{l_3,l'}_{\k+\q}}^*U^{l_4l}_\k{U^{l_2l}_\k}^*+U^{l_1,l'}_{\k+\q}{U^{l_4+5,l'}_{\k+\q}}^*U^{l_3+5,l}_\k{U^{l_2l}_\k}^*\right],
\label{chi-approx}
\end{eqnarray}
with $\Gamma_{ll',\k\q}=\gamma$ for $\gamma\ll1$.

When $\gamma$ is as large as $\Delta$, however,
we have to check to what extent eq. (\ref{chi-approx}) is reliable.
Considering that the origin of the renormalization of $\chi^s$
is the quasiparticle damping $\gamma(E^{l}_\k)$ and $\gamma(E^{l'}_{\k+\q})$,
we introduce the following approximation:
\begin{eqnarray} 
\Gamma_{ll',\k\q}=b\cdot {\rm max}\{\gamma(E^{l}_\k),\gamma(E^{l'}_{\k+\q})\}
\label{Gam}
\end{eqnarray}
where $b\approx1$ is a fitting parameter.
$\Gamma_{ll',\k\q}\approx0$ in the SC state for $|E^{l}_\k|,|E^{l'}_{\k+\q}|<3\Delta$,
reflecting the absence of quasiparticle damping.
In Fig. \ref{Fig1}, we show numerical results 
given by the present approximation with $b=1.3$;
we replace $b\gamma_0$ with $\gamma_0$ hereafter since $b\approx1$.
Since the ``exact results'' given by eq. 
(\ref{chi-exact}) is quantitatively reproduced,
we decide to calculate Im$\chi^s(\w,{\bf Q})$
using eqs. (\ref{chi-approx}) and (\ref{Gam})
for more realistic values of $\Delta$ and $\gamma$.
We verified that the present approximation works well
when $\gamma$ is comparable to or smaller than $\Delta$.

\begin{figure}[!htb]
\includegraphics[width=\linewidth]{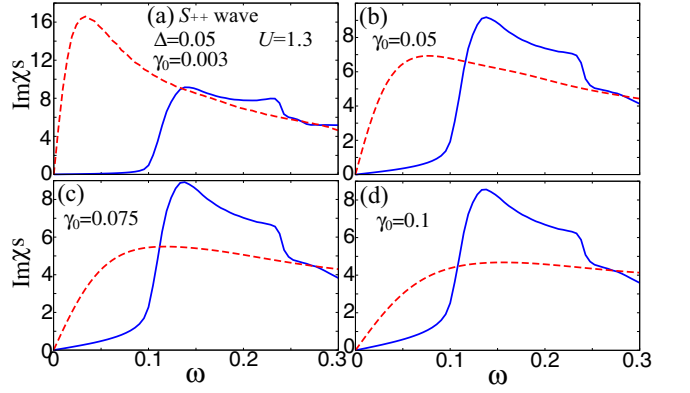}
\caption{
(Color online) Im$\chi^s(\w,{\bm Q})$ 
for $s_{++}$-wave (solid line) and normal (broken line) states,
with $\gamma_0=0\sim0.1$.
}
\label{Fig2}
\end{figure}

Figure \ref{Fig2} shows Im$\chi^s(\w,\Q)$ obtained by 
eqs. (\ref{chi-approx}) and (\ref{RPA}) for $U=1.3$ and $T=0.002$.
In the $s_{++}$-wave SC state, we put $\Delta=0.05$;
although it is a few times larger than the gap for Sm1111 with $T_{\rm c}=56$K,
it is enough smaller than the Fermi energies of electron- and hole-pockets
\cite{Kuroki}.
In the numerical calculation, we use $1024\times1024$ $\k$-meshes.
Hereafter, we put $a(3\Delta)=0.003/\gamma_0$.
When (a) $\gamma_0=0.003$,
Im$\chi^s(\w,{\bm Q})$ in the SC state approximately equal to that 
in the normal state for $\w>2\Delta$.
As $\gamma_0$ increases from (b) $0.05$ to (d) $0.1$,
Im$\chi^s$ in the normal state decreases gradually,
whereas that in the SC state depends on $\gamma_0$ only slightly,
since $\gamma(\e)\approx0$ for $|\e|<3\Delta$.
Therefore, in the case of $\gamma_0\gtrsim\Delta$,
Im$\chi^s(\w,\Q)$ in the SC state shows a prominent hump structure, 
and its peak value is about double of that in the normal state.
In (d), experimental approximate ``sum-rule'' at fixed $\q=\Q$
 \cite{keimer} is well satisfied.
In Fig. \ref{Fig2} (c) and (d),
a relatively large slope for $|\e|<2\Delta$
is an artifact of the approximation due to large $\gamma_0/\Delta$.





\begin{figure}[!htb]
\includegraphics[width=\linewidth]{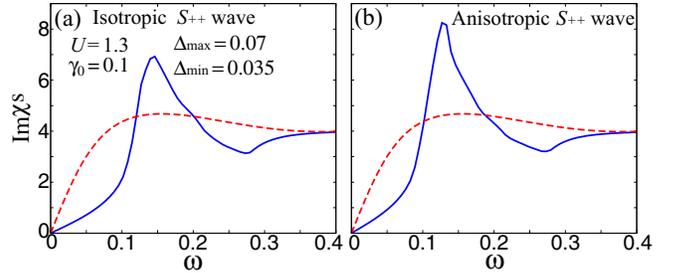}
\caption{
(Color online) Im$\chi^s(\w,{\bm Q})$ for $s_{++}$-wave (solid line)
and normal (broken line) states,
with $\Delta_{\rm max}=0.07$ and $\Delta_{\rm min}=0.035$.
}
\label{Fig3}
\end{figure}

Next, we study the effect of band-dependent SC gap
observed by ARPES measurements \cite{ARPES1,ARPES4}.
In Fig. \ref{Fig3} (a), we put $U=1.3$, 
$\Delta_{1,2,4}=\Delta_{\rm max}=0.07$eV for FS1,3,4, and 
$\Delta_2=\Delta_{\rm min}=0.035$eV for FS2.
In (b), we introduce the anisotropy of the gap function for only FS3,4
with ratio 2; $\Delta_\k=\Delta_{\rm max}(1-0.5\sin^2\theta_\k)$,
where $\theta_\k= \tan^{-1}(|k_{y(x)}|/(|k_{x(y)}|-\pi))$ for FS3(4).
Here, we put $a(\e)$ in eq. (\ref{a}) as
(i) $0.003/\gamma_0$ for $|\e|<3\Delta_{\rm min}$,
(ii) $1$ for $|\e|>4\Delta_{\rm min}$, and
(iii) linear extrapolation for $3\Delta_{\rm min}<|\e|<4\Delta_{\rm min}$.
In Fig. \ref{Fig3} (a), Im$\chi^s(\w,{\bm Q})$ increases
rapidly at $\w=\Delta_{\rm max}+\Delta_{\rm min}=0.105$, and 
it shows a peak at $\w=0.14$.
In (b), the peak is located at $\w=0.125$, which is 
closer to $\Delta_{\rm max}+\Delta_{\rm min}=0.105$.
In Fig. \ref{Fig3} (a) and (b),
the width of the hump peak is much sharper than that
for the band-independent SC gap in Fig. \ref{Fig2},
since Im$\chi^s(\w,{\bm Q})$ is reduced by damping
for $|\w|>4\Delta_{\rm min}=0.14$.
We have also calculated Im$\chi^s(\w,{\bm Q})$
for $\Delta_{3,4}=\Delta_{\rm max}$ and $\Delta_{1,2}=\Delta_{\rm min}$,
and verified that the obtained result is similar to Fig. \ref{Fig3}.

Here, we make comparison with experiments.
The peak height and the weight in Fig. \ref{Fig3} (b) seems to be 
consistent with the neutron scattering measurements in iron pnictides
\cite{christianson,keimer,zhao,qiu}.
In BaFe$_{1.85}$Co$_{0.15}$As$_2$ ($T_{\rm c}=25$K),
the observed "resonance energy'' is $\w_{\rm res}=9.5$ meV  \cite{keimer}.
According to ref. \cite{ARPES1}, $\Delta_{\rm max}/T_{\rm c}\approx 3.5$ and 
$\Delta_{\rm min}/\Delta_{\rm max}\approx 0.35$ in many iron pnictides.
(More smaller $\Delta_{\rm max,min}$ is reported in ref. \cite{Matsuda}.)
Thus, $\Delta_{\rm max}+\Delta_{\rm min}\approx 4.7T_{\rm c}=10$ meV
is comparable to $\w_{\rm res}=9.5$ meV in BaFe$_{1.85}$Co$_{0.15}$As$_2$.
Moreover, finite Im$\chi^s(\w,{\bm Q})$
for $\w\gtrsim0.3\w_{\rm res}$ in ref. \cite{keimer}
may suggest the existence of SC gap anisotropy.
Therefore, the theoretical result in Fig. \ref{Fig3} (b)
is well consistent with experimental data.
We have verified that the hump structure of Im$\chi^s(\w,\q)$ 
with $\q=(\pi,0)$ is very small for $\gamma_0\sim0.1$.

\begin{figure}[!htb]
\includegraphics[width=\linewidth]{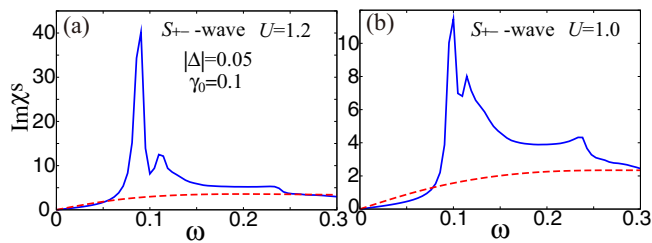}
\caption{
(Color online) Im$\chi^s(\w,{\bm Q})$ for $s_\pm$-wave (solid line)
and normal (broken line) states, with $U=1.2$ and $1.0$.
}
\label{Fig4}
\end{figure}

We also analyze the resonance peak for the $s_\pm$-wave state in
Fig. \ref{Fig4}.
In this case, the spin wave without damping is observed as the
``resonance peak'' at $\omega_{\rm res}<2\Delta$.
Figure \ref{Fig4} shows the numerical results 
for (a) $U=1.2$ and (b) $U=1.0$ in the case of $\Delta_{1,2}=-\Delta_{3,4}=0.05$.
In (a), a very sharp and high resonance peak appears 
at $\w_{\rm res}=0.85<2\Delta$, consistently with previous theoretical studies 
\cite{maier-scalapino,eremin}.
The case (b) with $U=1.0$ corresponds to the "heavily overdoped'' 
since $\a_{\rm St}=0.69$ and $T_{\rm c}\sim 0$.
The obtained resonance peak in Fig. \ref{Fig4} 
by taking $\g(\e)$ into account is too large to explain experiments even
in the case of $\a_{\rm St}=0.69$.
In Bi-based high-$T_{\rm c}$ cuprates, the width of the resonance peak is wide 
due to the sample inhomogeneity (i.e., nanoscale distribution of $T_{\rm c}$)
 \cite{keimer-highTc}.
However, weight of the peak is 10 times larger than that in 
BaFe$_{1.85}$Co$_{0.15}$As$_2$ \cite{keimer}.


In the present study, we have neglected the impurity effect
since its influence on $\chi^s(\w,\Q)$ is expected to be small.
In fact, in the single band model, the reduction in $\chi^0$ due to the 
impurity self-energy is almost canceled by the impurity vertex correction
 \cite{Bulut}.
Moreover, impurity effect tends to {\it enhance} $\chi^s(\w,\Q)$
in the modified FLEX approximation in nearly AF metals \cite{ROP}.

Before closing the study, we shortly discuss 
the heavy fermion Kondo insulator CeNiSn.
As shown in Fig. 1 of Ref. \cite{CeNiSn},
neutron scattering spectrum at $\q=(0,\pi,0)$ in CeNiSn 
shows a prominent hump peak structure above the hybridization gap below 
the Kondo temperature $T_{\rm K}$, which looks very similar to 
the spectrum observed in iron pnictides below $T_{\rm c}$
\cite{christianson,keimer,zhao,qiu}.
This hump structure is well reproduced by the dynamical-mean-field theory 
based on the periodic Anderson model \cite{mutou-hirashima}.
This fact demonstrates that large hump in Im$\chi^s(\w,\Q)$
can appear in strongly correlated systems with one-particle gap,
without the necessity of the resonance mechanism.

In summary, we have studied Im$\chi^s(\w,\Q)$ in iron pnictides
based on the five-orbital model,
and revealed that a prominent hump structure appears 
just above $2\Delta$ in the $s_{++}$-wave state,
by taking the strongly correlation effect via $\gamma$.
This hump structure becomes small as $\alpha_s$ decreases 
in the over-doped region or $q$ deviates from the nesting.
At present, experimental data can be explained 
in terms of the $s_{++}$-wave state very well.
Further experimental efforts are required to determine 
the height and width of the "resonance peak'', and 
the magnitude relation between $\w_{\rm res}$ and
$\Delta_{\rm max}+\Delta_{\rm min}$.

\acknowledgements
This study has been supported by Grants-in-Aid for Scientific 
Research from MEXT of Japan, and by JST, TRIP.
Numerical calculations were performed using the facilities of 
the supercomputer center, ISSP.

\end{document}